\documentclass[12pt,preprint]{aastex}
\usepackage{natbib,graphicx}

\def\wig#1{\mathrel{\hbox{\hbox to 0pt{%
          \lower.5ex\hbox{$\sim$}\hss}\raise.4ex\hbox{$#1$}}}}

\shorttitle{Stellar Carbon-to-Oxygen Ratios}
\shortauthors{Fortney}

\newcommand{\ms}{$M_{\odot}$}

\newcommand{\te}{$T_{\rm eff}$}
\newcommand{\teff}{$T_{\rm eff}$}

\newcommand{\cp}{\citep}
\newcommand{\ct}{\citet}

\slugcomment{ApJ}

\begin{document}

\title{On the Carbon-to-Oxygen Ratio Measurement in Nearby Sunlike Stars:  Implications for Planet Formation and the Determination of Stellar Abundances}

\author{Jonathan J. Fortney\altaffilmark{1}}

\affil{Department of Astronomy and Astrophysics, University of California, Santa Cruz, CA 95064; jfortney@ucolick.org}
\altaffiltext{1}{Alfred P. Sloan Research Fellow} 

\begin{abstract}
Recent high resolution spectroscopic analysis of nearby FGK stars suggests that a high C/O ratio of greater than 0.8, or even 1.0, is relatively common.  Two published catalogs find C/O$>0.8$ in 25-30\% of systems, and C/O$>1.0$ in $\sim$~6-10\%.  It has been suggested that in protoplanetary disks with C/O$>0.8$ that the condensation pathways to refractory solids will differ from what occurred in our solar system, where C/O$=0.55$.  The carbon-rich disks are calculated to make carbon-dominated rocky planets, rather than oxygen-dominated ones.  Here we suggest that the derived stellar C/O ratios are overestimated.  One constraint on the frequency of high C/O is the relative paucity of carbon dwarfs stars ($10^{-3}-10^{-5}$) found in large samples of low mass stars.  We suggest reasons for this overestimation, including a high C/O ratio for the solar atmosphere model used for differential abundance analysis, the treatment of a Ni blend that affects the O abundance, and limitations of one-dimensional LTE stellar atmosphere models.  Furthermore, from the estimated errors on the measured stellar C/O ratios, we find that the significance of the high C/O tail is weakened, with a true measured fraction of C/O$>0.8$ in 10-15\% of stars, and C/O$>1.0$ in 1-5\%, although these are still likely overestimates.  We suggest that infrared T-dwarf spectra could show how common high C/O is in the stellar neighborhood, as the chemistry and spectra of such objects would differ compared to those with solar-like abundances.  While possible at C/O$>0.8$, we expect that carbon-dominated rocky planets are rarer than others have suggested.
\end{abstract}

 
\section{Introduction}
\subsection{The Composition of Stars and Planets}
The determination of the abundances of atoms in the atmospheres of stars is an essential element of modern astronomy.  Recently, tremendous work has occurred on understanding the relationship between planets and the abundances of planet-hosting and non-planet-hosting stars.  Since the pioneering work of \ct{Gonzalez97}, many investigators have worked to understand connections between stellar abundances and the observed frequency \cp{Santos04,Fischer05,Johnson10} and composition \cp{Guillot06,Burrows07,Miller11} of planets.

Our solar system is one realization of the complex planet formation process.  The raw materials that made up the Sun and solar nebula, through a process of condensation, grain growth, and accumulation, gave rise to four rocky planets in our inner solar system that are predominantly composed of Mg-Si-O-bearing rocks and Fe-Ni metals.  In other solar systems, with parent star disks with other abundances, a different selection of refractory materials, or in different relative abundances, surely occur.  For instance, if a nebula's carbon-to-oxygen (C/O) ratio is $\gtrsim$0.8, condensation pathways can change dramatically, leading to carbon-dominated rocky planets, as recently discussed in detail by \ct{Bond10}.

There had been prior intermittent interested in carbon-dominated planets in the past decade, from \ct{Gaidos00}, \ct{Lodders04}, and \ct{Kuchner05}, to name three examples.  In particular, \ct{Gaidos00} discussed different formation scenarios for giant planet cores and rocky planets in disks with varied C/O ratios, as well as how the chemical evolution of the galaxy generally can lead to enhanced C/O through time.  \ct{Lodders04} suggested that the planetesimals that make up Jupiter's heavy element enrichment were carbon-rich, and that Jupiter initially formed at the ``tar line'' rather than the ``ice line.''  This is one possible explanation for the low water abundance measured by the \emph{Galileo Entry Probe} \cp{Wong04}.  \ct{Kuchner05}, similar to \ct{Gaidos00}, were interested in giant planets and terrestrial planets that could form in environments where the local (or entire disk's) C/O$>1$, leading to carbon-dominated (rather than oxygen-dominated) silicates.

More recently \ct{Bond10} coupled protoplanetary disk abundances derived from stellar spectra to a model of disk chemistry, which yields the condensation sequence of solids.  Their work further coupled the formation of solids to an N-body model of planet formation \cp{OBrien06}.  For particular planetary systems, with measured C/O and Mg/Si ratios of the host star, they calculated the equilibrium disk chemistry and solid composition for the initial planetesimal distribution.  \ct{Bond10} furthermore kept track of the contribution of particular planetesimals as they add their mass to growing protoplanets, and in the end find the relative contributions of C, O, Mg, Si, etc.~to the masses of formed planets.

Within the context of giant planets, \ct{Madhu11} suggested that day-side photometry of the transiting planet WASP-12b indicates an atmosphere with C/O$>1.0$.  More recently \ct{Madhu11b} and \ct{Oberg11} have investigated the accumulation of gas and icy planetesimals in disks with a range of C/O ratios to understand possible pathways to forming ``carbon-rich'' gas giants.

\subsection{C/O Ratio in Stars}
Composition-dependent planet formation models depend on the stellar abundances of C and O for the initial conditions of disk chemistry.  The stellar C/O ratios in \ct{Bond10} were taken from determinations of C from \ct{Ecuvillon04} and of O from \ct{Ecuvillon06}.  Motivated by \ct{Bond10}, larger tabulations of C and O abundances were recently made by \ct{Delgado10}, for 370 FGK stars from the HARPS planet-search sample, and \ct{Petigura11}, for 457 F and G stars from the California Planet Survey sample.  These two studies are relatively similar, as they cover large samples that include planet-hosting stars and those not found to host planets.

The two studies do have some differences in the lines of C and O chosen.  For the carbon abundance, the \ct{Delgado10} work used CI lines at 5380.3 and 5052.2 \AA, with only 5380.3 \AA\ used for stars with \teff$< 5100$ K.  For oxygen, the forbidden lines of [OI] at 6300 and 6363 \AA\ were used.  The derivation of the abundances was done with a combination of the code MOOG, for the generation of synthetic spectra \cp[][as updated in 2002]{Sneden73}, the Kurucz ATLAS9 atmosphere grid with overshooting \cp{Kurucz93}, and the equivalent widths were measured using the ARES program \cp{Sousa07}.  The \ct{Petigura11} study used a CI line for carbon at 6587 \AA, and the [OI] line for oxygen at 6300 \AA.  The derivation of the abundances was performed with the Spectroscopy Made Easy (SME) code \cp{Valenti96} with Kurucz stellar atmospheres.

\begin{figure}[ht!] 
\begin{center}
 \includegraphics[width=4.5in]{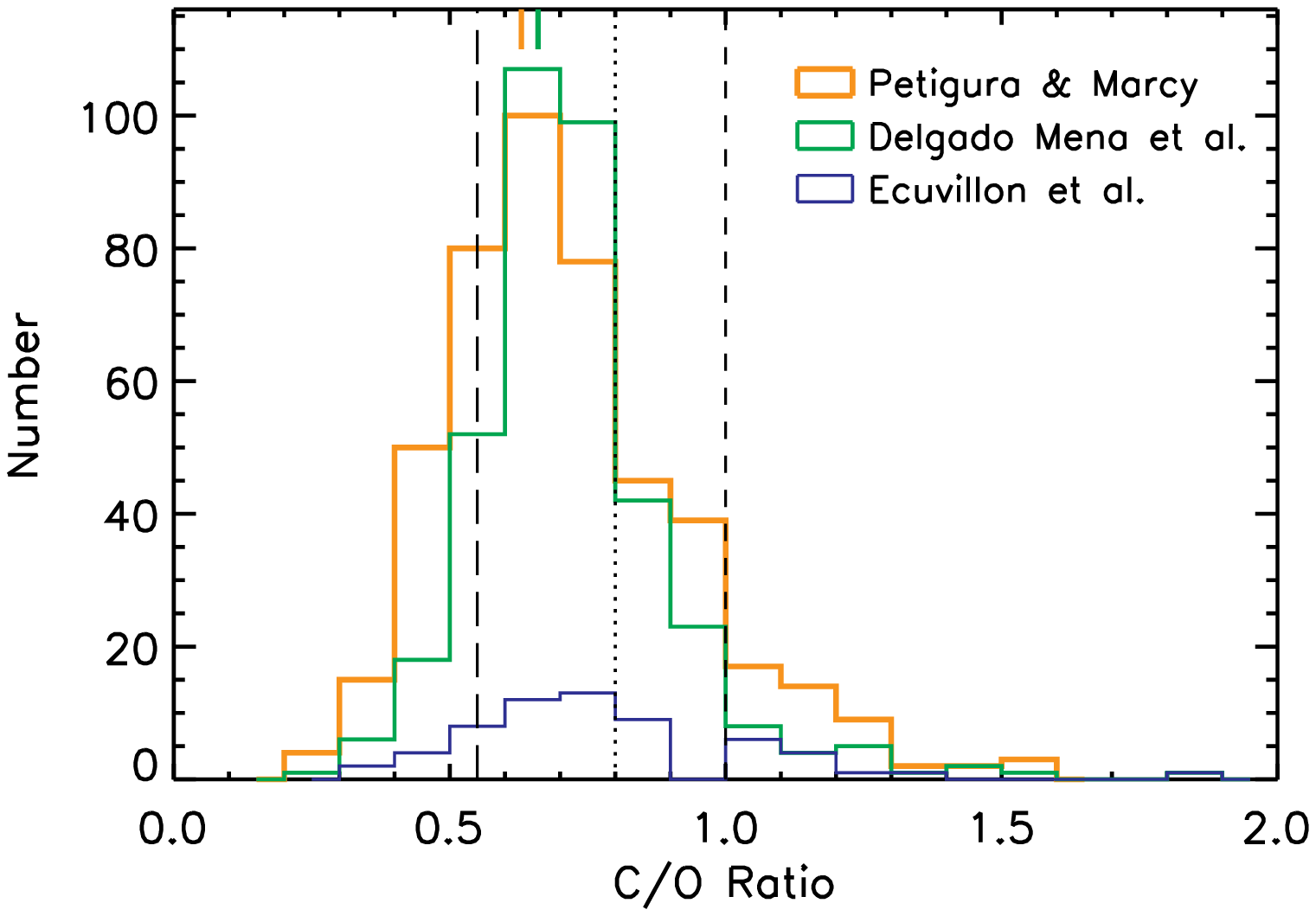}
 \includegraphics[width=4.5in]{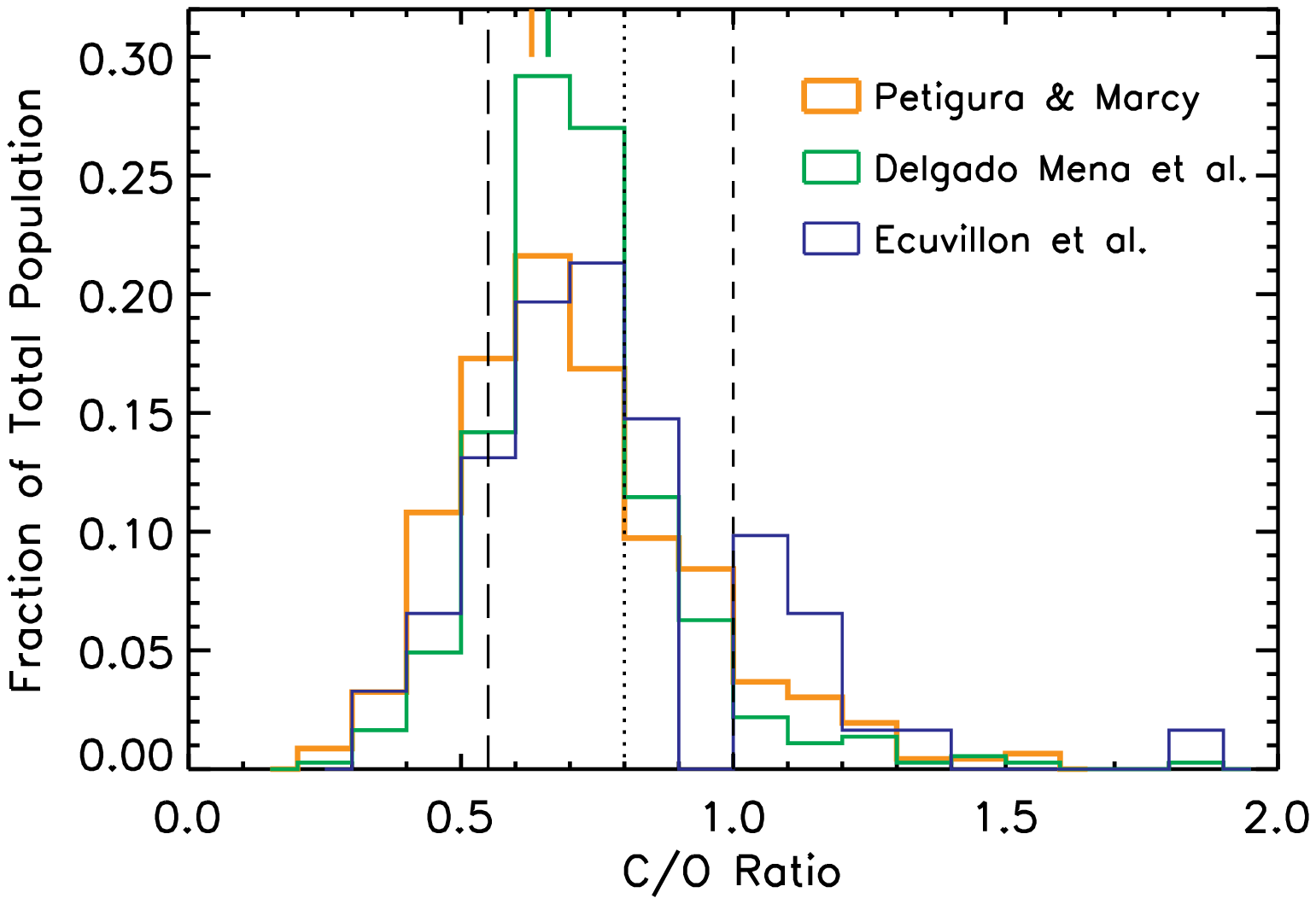}
 \caption{\emph{Top}: Histogram of the C/O ratios from the papers of \ct{Ecuvillon04,Ecuvillon06} (as tabulated in \citealp{Bond10}), \ct{Delgado10}, and \ct{Petigura11}, in blue, green, and orange, respectively.  C/O ratios of 0.55 (the solar value from Asplund et al.~2009), 0.8, and 1.0 are shown as long dashed, dotted, and short dashed lines, respectively.  The adopted solar C/O ratios from \ct{Delgado10} and \ct{Petigura11} are shown as thick green and orange ticks at the top. \emph{Bottom}:  The data sets are normalized, yielding fractions of each sample, instead of number.  The large samples of \ct{Delgado10} and \ct{Petigura11} find 24-29\% of FGK stars have C/O$>0.8$, and 6-10\% have C/O$>1$.}
   \label{histo1}
\end{center}
\end{figure}

Tabulations from \ct{Bond10} (who quote Ecuvillon et al.~values), \ct{Delgado10}, and \ct{Petigura11} are shown in \mbox{Figure \ref{histo1}}.  Both of the large studies found a somewhat similar shape.  They found a maximum at C/O ratios modestly higher that that of the Sun \cp[0.55, from][]{Asplund09}, with a noticeably enhanced peak in the distribution found by \ct{Delgado10}, shown most clearly in \mbox{Figure \ref{histo1}}\emph{b}.  Of particular interest to all of these authors, and our \emph{Letter}, is the tail off to higher C/O ratios $>0.8$ (dotted line) and even further onto $>1.0$ (short dashed line).  \ct{Delgado10} find C/O$>0.8$ for 24\% of their stars, and C/O$>1.0$ for 6\%.  For the \ct{Petigura11} sample, they find C/O$>0.8$ for 29\% of their stars, and C/O$>1.0$ for 10\%.  The numbers quoted are for the mixed sample of planet-hosting and non-planet hosting stars.  Taken at face value, and the condensation chemistry in \ct{Bond10}, this potentially implies carbon planets (formed when C/O$>0.8$) in $\sim$25\% of planetary systems.
 
However, one must take care when estimating the fraction of stars with high C/O ratios, given observational error bars.  Of interest is the positive tail at high C/O, and a tail such as this is expected since the error is approximately constant in the logarithmic abundance ratio, [C/O].  This leads to an error distribution that is log-normal in the C/O ratio, and is seen in, for instance, Table 6 in \ct{Petigura11}.  Their average error at C/O$=1$ is 0.23, which is 1.61$\sigma$ above the solar C/O ratio they use, of 0.63.  From this 1.61$\sigma$, 5.4\% of the stellar sample is thus expected to be found with C/O$>1$, just due to observational errors.  This would yield a ``true'' fraction of stars with C/O$>1$ of $\sim$5\%, rather than 10\%.  \ct{Delgado10} do not provide individual errors on their C/O determinations, but assuming similar errors, their entire sample of C/O$>1$ stars (6\%) can be explained by errors.  At C/O$=0.8$, in the \ct{Petigura11} sample, the average error is 0.16, a difference of 0.17 above their solar value, or 1.06$\sigma$.  This yields an expected fraction of 14.5\%, which would move the 29\% with C/O$>0.8$ down to 15\%.  For \ct{Delgado10}, this expected fraction moves their 24\% at C/O$>0.8$ down to 10\%.  The upshot is a much reduced tail of stars with high C/O ratios.

\section{Motivation}

Deriving the abundances of C and O in stellar atmospheres is a demanding task.  Even for the Sun, this has been an especially difficult over the past decade \cp[e.g.,][]{Allende01,Allende02,Asplund09}.  The most recent state-of-the-art work includes intense efforts to find unblended lines, calculated NLTE corrections where applicable, and solar atmosphere pressure-temperature profiles that come directly from three-dimensional simulations \cp{Asplund09,Caffau11}.

While deriving the abundances of atoms and molecules in low-mass M and K dwarf stellar atmospheres is more difficult that than of Sunlike stars, they do have one natural asset.  Cooler dwarf atmospheres have a natural ``flip'' in the chemistry at C/O$=1$.  At C/O$<1$, most O goes into the CO molecule, but additional O is left over and partitions into H$_2$O and, for M stars, TiO/VO gases.  TiO/VO absorption bands dominate the optical spectra of M stars.  However, at C/O$>1$, the CO molecules use up nearly all O in cool stars, leaving excess C, which goes into molecules such as C$_2$ and CN.  In such a cool star, the optical region is dominated by C$_2$ and CN.  We suggest that if C/O$>1$ in FGK stars is a relatively common phenomenon then cool carbon stars showing C$_2$ and CN bands should be relatively common in the solar neighborhood.  This is something that can be investigated, and was even briefly noted by \ct{Petigura11}.

\section{True Fraction of Carbon Dwarfs?}
Carbon dwarfs, which appear to have C/O$>1$ in their atmosphere, are rare stars.  The best current understanding of their formation is that they are \emph{not} made of primordial C/O$>1$ gas, but instead have high C abundances due to accretion of AGB dredge-up material from a companion \cp[e.g.,][]{deKool95,Steinhardt05}.  Carbon dwarfs have been found by many authors, with most being found by large surveys, such as the SDSS \cp{Margon02,Downes04}.  Their numbers are predominantly spread among what would otherwise be G, K, and M spectral types.

For our purposes here, instead of forming these carbon-rich stars by accretion, we can take the extremely optimistic view that these are in fact primordial C/O$>1$ systems.  Then the fraction of dwarfs that are carbon dwarfs would be the upper limit on the fraction of stars with primordial C/O$>1$.  \ct{deKool95} have previously estimated the frequency of carbon dwarfs, based on detections before SDSS.  They estimate a space density in the disk of $\sim 1 \times 10^{-6}$ pc${-3}$, but noted this may be an overestimate.  This can be compared to the space density of stars with mass $<0.7$ \ms, $6.5 \times 10^{-2}$ pc${-3}$, from \ct{Bochanski10}.  This is a 4-5 order of magnitude difference.

There are other ways to estimate the carbon dwarf frequency.  In particular, \ct{Covey08} have analyzed SDSS spectra and 2MASS photometry of 25,000 sources, in an effort to take a census of low mass stars out to $J$=16.2, over a mass range from 0.1 - 0.7 \ms.  For these cool stars, if C/O$>1$ did occur, it should be clear, given the abundance of carbon-rich molecules (C$_2$, CN) that would occur in such cool atmospheres.  Of their sample of 9,649 low mass stars, they note 24 ``exotic contaminants'' including 4 carbon stars.  This implies a C/O$>1$ in only 0.04\% of cases, although based on color cuts $\sim$1/2 of the carbon dwarfs may have been missed (K.Covey, personal communication).  This work, and the fact that \ct{Downes04} found only $\sim$100 carbon dwarfs from a large SDSS search at $15.6<r<20.8$ to specifically find such stars, certainly strengthens the point of \ct{deKool95} that carbon dwarfs are quite rare.  The relative frequency is likely $10^{-3}-10^{-5}$.

We can think of no reason that Sunlike stars in the solar neighborhood would be uniformly more enriched in carbon (or oxygen deficient) than the KM stars.  One potential way out could be if the later-type stars are systematically older than the earlier-type stars, so that some amount of galactic chemical evolution could have taken place.  In this case the younger (FG) stars could have higher C/O \cp[e.g.][]{Gaidos00}.  However, the HARPS/California samples are not young stars, so constructing a credible explanation through this path seems unlikely.

\section{Diagnosing high C/O from Brown Dwarfs?}
The K \& M dwarfs from the SDSS give us leverage into the fraction of dwarfs with C/O$>1$.  Are there are any stellar populations that could be used to understand the fraction of stars with $0.8<$C/O$<1$?  One possibility are the L and T dwarfs.  L-type dwarfs, from \te s of $\sim$~2400-1400 K, have atmospheres whose infrared opacity is dominated by molecular bands of H$_2$O, CO, and cloud layers made up of refractory condensates like corundum (Al$_2$O$_3$), enstatite (MgSiO$_3$), forsterite (Mg$_2$SiO$_4$), and iron (Fe) \cp{AM01}.  Corundum, enstatite, and forsterite need abundant oxygen to form.  If C/O$>0.8$, or C/O$>1$, are common in L-dwarfs, then perhaps we could see a different kind of cloudy L-dwarf, with condensates dominated by SiC instead of the Mg-Si-O forms.  Certainly the spectra of such objects would be highly abnormal at C/O$>1$, as they would strongly favor CO at the expense of H$_2$O, but perhaps even at C/O ratios closer to 0.8, the cloud condensation pathway would be different than normal L dwarfs. 

In the cooler objects, the T-dwarfs, the silicate clouds appear to reside below most of the visible atmosphere.  Quite importantly, these refractory clouds remove 21\% of oxygen from the atmosphere \cp[e.g.][and references therein]{Visscher10}.  This removal of oxygen from the gas raises the C/O ratio in the visible atmosphere.  \ct{Lodders10} have investigated chemistry as a function of C/O ratio at 1100 K and 0.01 bar, which is after silicate condensation.  Lodders finds a dramatic change in chemistry even at C/O$=0.8$.  At C/O$>0.8$, methane becomes more abundant than water (by two orders of magnitude at C/O$=0.9$) and HCN becomes nearly as abundant as water.  Since CO, CH$_4$, H$_2$O (and HCN) all have prominent opacity in the infrared, C/O$>0.8$ brown dwarfs could appear distinctly different than those with a C/O ratio closer to that of the Sun.

Since most brown dwarfs are found from color-color diagrams in surveys such as SDSS and 2MASS, it is quite possible that high C/O brown dwarfs, should they exist, could have previously eluded detection.  The cuts from SDSS use $i-z$ color, which could in principle be less sensitive to the C/O chemistry since the optical spectra of brown dwarfs are shaped by pressure-broadened lines of Na and K \cp{bms}.  But 2MASS uses $JHK_{\rm s}$ cuts, which could lead to high C/O dwarfs being missed, since in some cases H$_2$O would not be the dominant opacity source, leading to different near-infrared colors.  As far as we are aware, there are no compelling outliers for high C/O brown dwarfs, but a search for them could be valuable.

\begin{figure}\epsscale{0.75}
\plotone{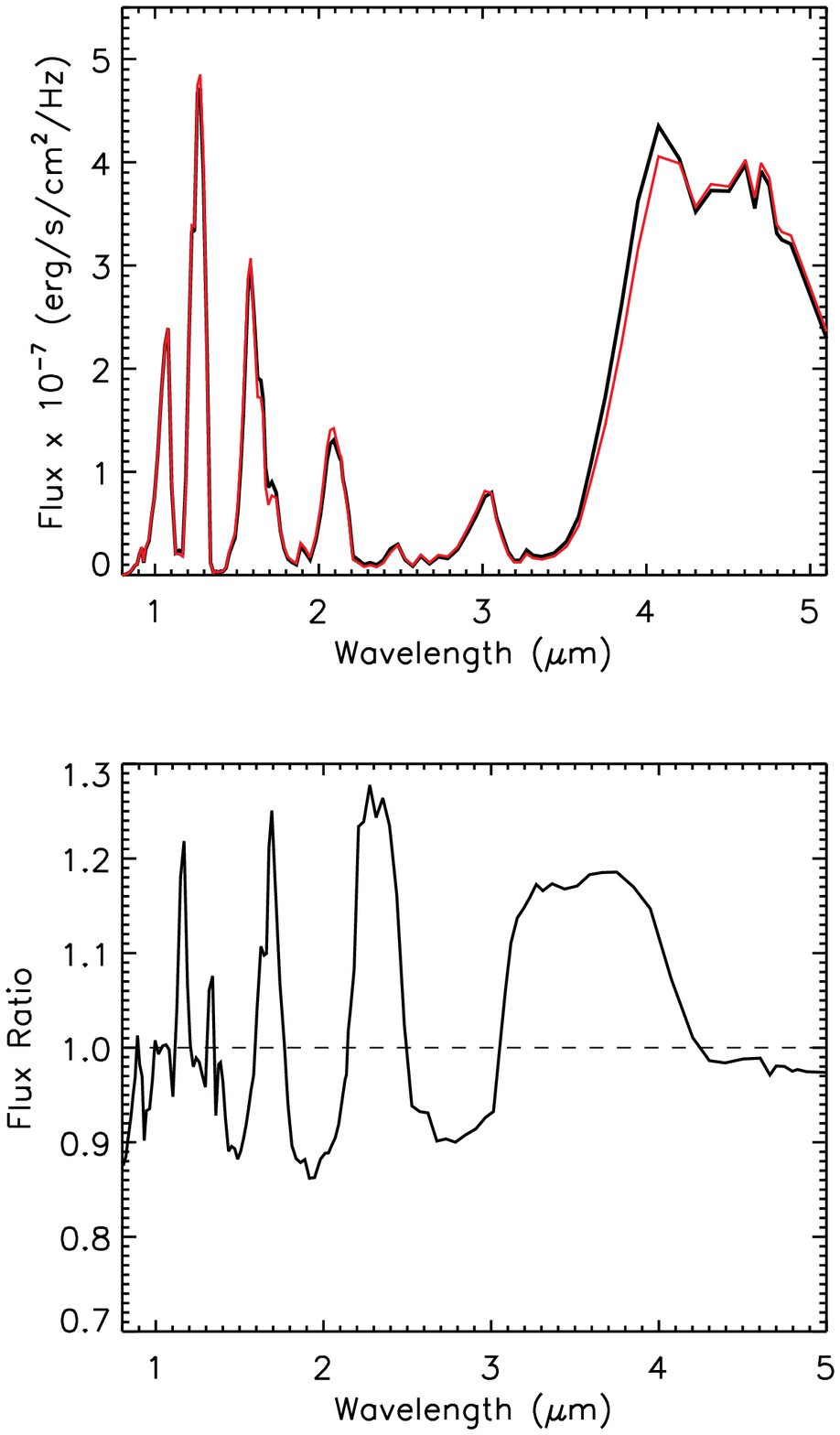}
\caption{\emph{Top}: Emitted spectra from two models of a brown dwarf with \te $=900$K and log $g=5.0$.  The thick black curve is for solar abundances, at C/O$=0.55$.  The thin red curve is for a model with C/O$=0.7$, with the carbon abundance enhanced. \emph{Bottom}:  The ratio of the flux between these two models.
\label{ratio}}
\end{figure}

A full exploration of C/O ratio in brown dwarfs, and its affects on clouds, is beyond the scope of this work.  However, there is an additional point that we can motivate.  In T dwarfs, essentially all infrared opacity is due to water, methane, and carbon dioxide, which tie up nearly all C and O.  The relatively cloud-free spectra of T dwarfs may allow for accurate determinations of C/O in the solar neighborhood.  In Figure \ref{ratio} we show model spectra of two 900 K dwarfs, one with C/O$=0.55$ (the solar value) and another at C/0$=0.7$.  These models were computed using the atmosphere code of M.~Marley and J.~Fortney and collaborators, described elsewhere \cp{Marley02,Fortney07b}.  The opacity and chemistry databases are taken from a previous tabulation \cp{Fortney05}.  The differences between the two spectra can be prominent, in particular as a ratio.  One can see differences of 10\% in the JHK peaks, and 20-30\% in the water bands.  With future improvements in the accuracy of opacities, in particular a high-temperature database for CH$_4$, we suggest it may be possible to derive relative C and O abundances, from H$_2$O, CH$_4$, and CO, from infrared T-dwarf spectra.  High spectral resolution observations of brown dwarfs are now being achieved as well \cp{Rice10}, which could allow further progress in abundance analysis.  The utility of such efforts would be a better understanding of the C/O ratio of stars in the solar neighborhood.  Recently, some work by \ct{Tsuji11} on a small number of brown dwarfs with AKARI observations has moved in this direction.

\section{Discussion}
We suggest that recent high-resolution spectroscopic analyses have overestimated the fraction of FGK stars with C/O$>1.0$, and, by extension, perhaps with C/O$>0.8$ as well.  \ct{Gaidos00} and \ct{Bond10} suggested that the pathway to form ``rocky'' planets in C/O$>0.8$ systems will differ in these carbon-rich system, making carbon-rich planets.  While this logic seems secure, we find it is less likely than some have anticipated.  This is because the true fraction of carbon-rich parent stars is quite low, $10^{-3}-10^{-5}$.  However, our work should not be interpreted as claiming that carbon-rich terrestrial or giant planets cannot form.  The C/O ratio surely varies in the ISM and the region of phase space between $0.8<\mathrm{C/O}<1$ is less constrained by our work.  However, given the overestimation at C/O$>1$ it appears probable that this region is less-populated than has been recently suggested \cp{Bond10,Delgado10,Petigura11}, although not empty.  Furthermore we showed in \S1.2 that the quoted observational error bars from these groups imply smaller measured fractions of high C/O stars, which eliminated 15\% from C/O$>0.8$ and 5\% from C/O$>1.0$.

These authors of course put in great effort to understand the sources of their errors, and to correct for them.  The choice of lines used could be re-examined.  The choice of low-excitation forbidden [OI] lines along with high excitation permitted CI lines could introduce systemic errors.  A revised (although difficult) study could examine [CI]/[OI] or CI/OI, since these lines would behave similarly at a given stellar temperature and in deviations from LTE  (M.~Asplund, personal communication).

An additional point is that both the large \ct{Delgado10} and \ct{Petigura11} works tuned their abundance retrieval methods to match Sun's C/O ratio, as a standard.  \ct{Asplund09}, in a recent review, find C/O$=0.55$.  However, in \ct{Delgado10} the particular abundances used for the Sun lead to C/O$=0.66$, while in \ct{Petigura11}, it is 0.63.  (See the thick tick marks at the top of Figure 1).  So there clearly could be an offset of $\sim$0.1, towards higher C/O, in these works.

An issue with the Ni blend of the [OI] line used by \ct{Petigura11} may potentially skew their results towards higher C/O ratios.  These authors adopt the abundances of O and Ni from 3D abundance analyses, and then use them in their 1D analysis.  To obtain a good fit to the solar spectrum, they adopted $\mathrm{log}(gf)=-1.98$ for the NiI line, which is 35\% higher than the laboratory measurement of \mbox{-2.11} measured by \ct{Johansson03}.  Consequently, \ct{Petigura11} in their Figure 3, show a NiI line that is 40\% stronger than found in, for example, \ct{Allende01}.  This could explain the upward trend in C/O with increasing [Fe/H] in their Figure 16, as follows:  a known trend is that [Ni/Fe] increases slightly \cp[e.g.][]{Robinson06} and [O/Fe] decreases as metallicity ([Fe/H]) increases \cp[e.g.][]{Delgado10}.  A Ni blend that is too strong in the solar fit will increase in importance as [Fe/H] increases, leading to an overestimation of the Ni blend, an underestimation of the O abundance, and a C/O ratio that is too high.  In the \ct{Petigura11} sample, this could explain why most of the C/O$>1$ stars occur when [Fe/H]$>0.2$.

We note that \ct{Delgado10} and \ct{Petigura11} both relied on the stellar atmosphere models of Kurucz.  As a comparison, one could alternatively use the PHOENIX \cp{Hauschildt99} or MARCS \cp{Gustafsson08} stellar atmospheres. \ct{Gustafsson08} have compared their MARCS models to published models from Kurucz and PHOENIX, and the agreement does indeed vary with \teff\ and with plane parallel vs.~spherical symmetry.

There are still other pathways towards understanding extrasolar abundances.  As FG stars evolve onto the red giant branch and cool to \teff\ values of $\sim$~4000 K, it may be easier to constrain their C/O ratios in the same qualitative manner we suggest for M dwarfs.  The composition of extrasolar planetesimals can be determined by studying externally polluted white dwarf atmospheres, and some early evidence has been found for carbon-poor planetesimals \cp{Jura06}.

While we have focused exclusively on stellar abundances, it is important to recall that condensation of solids within a disk itself can potentially lead to non-stellar C/O ratios in nebula gas.  For instance, condensation of water in a protoplanetary disk can leave the surrounding nebula gas relatively carbon-rich (through large abundance of gaseous CO), which can change the relative ratios compared to those of the parent star \cp{Stevenson88,Lodders10,Oberg11}.  This may be a pathway to forming giant planets with relatively high C/O ratios in their envelopes.  A giant planet atmosphere with C/O$>1$ was recently suggested for planet WASP-12b, based on fits to 7 photometric points of day-side planet emission \cp{Madhu11}.  Their finding could be made more robust with the inclusion into their model of the absorption bands of molecules that are expected in the high C/O regime, such as HCN and C$_2$H$_2$ \cp{Lodders10}, which were not considered in the \ct{Madhu11} study.

Carbon-dominated rocky planets would be extremely interesting objects.  Their prevalence around stars of Sunlike abundances, and those stars with enhanced C/O ratios, would certainly tell us much about nebular condensation chemistry and the planet formation process.  While these planets may be inherently rare, we look forward to additional advances in the future.

\acknowledgements
JJF acknowledges support from the Alfred P.~Sloan Foundation.  JJF thanks the anonymous referee for a important suggestions regarding statistics and nickel blends.  JJF  thanks Debra Fischer, Martin Asplund, Bruce Margon, Katharina Lodders, Channon Visscher, Mark Marley, Mike Irwin, Kevin Covey, Mike Jura, Graeme Smith, Greg Laughlin, Geoff Marcy, Travis Barman, Sean Raymond, Andrew Howard, Jade Bond, and David O'Brien for many stimulating conversations throughout the course of this project.  JJF thanks Jacob Bruns for compiling the C/O data.


\end{document}